\documentclass[pra,preprint,showpacs,superscriptaddress,floatfix]{revtex4}\voffset-.5truecm
\font\titolone=cmbx10 scaled\magstep 2%
\font\sc=cmcsc10%
\font\ottorm=cmr8%
\def\st{\scriptstyle}%
\def\dt{\displaystyle}%
\font\tenmib=cmmib10 \font\eightmib=cmmib8
\font\sevenmib=cmmib7\font\fivemib=cmmib5 
\font\ottoit=cmti8\font\fiveit=cmti5\font\sixit=cmti6%%
\font\fivei=cmmi5\font\sixi=cmmi6\font\ottoi=cmmi8
\font\ottorm=cmr8
\font\ottosy=cmsy8\font\sixsy=cmsy6\font\fivesy=cmsy5%%
\font\ottobf=cmbx8\font\sixbf=cmbx6\font\fivebf=cmbx5%
\font\ottocss=cmcsc8%
\def\ottopunti{\def\rm{\fam0\ottorm}\def\it{\fam6\ottoit}%
\def\bf{\fam7\ottobf}%
\textfont1=\ottoi\scriptfont1=\sixi\scriptscriptfont1=\fivei%
\textfont2=\ottosy\scriptfont2=\sixsy\scriptscriptfont2=\fivesy%
\textfont4=\ottocss\scriptfont4=\sc\scriptscriptfont4=\sc%
\textfont5=\eightmib\scriptfont5=\sevenmib\scriptscriptfont5=\fivemib%
\textfont6=\ottoit\scriptfont6=\sixit\scriptscriptfont6=\fiveit%
\textfont7=\ottobf\scriptfont7=\sixbf\scriptscriptfont7=\fivebf%
\setbox\strutbox=\hbox{\vrule height7pt depth2pt width0pt}%
\normalbaselineskip=9pt\rm}
\let\nota=\ottopunti%
\mathchardef\Ba   = "050B  %alfa
\mathchardef\Bb   = "050C  %beta
\mathchardef\Bg   = "050D  %gamma
\mathchardef\Bd   = "050E  %delta
\mathchardef\Be   = "0522  %varepsilon
\mathchardef\Bee  = "050F  %epsilon
\mathchardef\Bz   = "0510  %zeta
\mathchardef\Bh   = "0511  %eta
\mathchardef\Bthh = "0512  %teta
\mathchardef\Bth  = "0523  %varteta
\mathchardef\Bi   = "0513  %iota
\mathchardef\Bk   = "0514  %kappa
\mathchardef\Bl   = "0515  %lambda
\mathchardef\Bm   = "0516  %mu
\mathchardef\Bn   = "0517  %nu
\mathchardef\Bx   = "0518  %xi
\mathchardef\Bom  = "0530  %omi
\mathchardef\Bp   = "0519  %pi
\mathchardef\Br   = "0525  %ro
\mathchardef\Bro  = "051A  %varrho
\mathchardef\Bs   = "051B  %sigma
\mathchardef\Bsi  = "0526  %varsigma
\mathchardef\Bt   = "051C  %tau
\mathchardef\Bu   = "051D  %upsilon
\mathchardef\Bf   = "0527  %phi
\mathchardef\Bff  = "051E  %varphi
\mathchardef\Bch  = "051F  %chi
\mathchardef\Bps  = "0520  %psi
\mathchardef\Bo   = "0521  %omega
\mathchardef\Bome = "0524  %varomega
\mathchardef\BG   = "0500  %Gamma
\mathchardef\BD   = "0501  %Delta
\mathchardef\BTh  = "0502  %Theta
\mathchardef\BL   = "0503  %Lambda
\mathchardef\BX   = "0504  %Xi
\mathchardef\BP   = "0505  %Pi
\mathchardef\BS   = "0506  %Sigma
\mathchardef\BU   = "0507  %Upsilon
\mathchardef\BF   = "0508  %Fi
\mathchardef\BPs  = "0509  %Psi
\mathchardef\BO   = "050A  %Omega
\mathchardef\BDpr = "0540  %Dpr
\mathchardef\Bstl = "053F  %*
\def\fiat{}
\let\a=\alpha     \let\d=\delta  \let\e=\varepsilon
  \let\h=\eta   \let\th=\theta \let\k=\kappa  \let\l=\lambda
\let\m=\mu                  \let\r=\rho
\let\s=\sigma \let\t=\tau    
      
 \let\D=\Delta

\def\CC{{\cal C}}

\let\ig=\int
\let\io=\infty
\let\==\equiv
\let\0=\noindent

\def\\{\hfill\break}

\def\*{\vskip3mm} 
\let\dpr=\partial
\def\defi{\,{\buildrel def\over=}\,}
\def\V#1{{\underline#1}}
\def\media#1{{\langle#1\rangle}}
\def\fra#1#2{{#1\over#2}}
\def\crcl{\,\raise.5mm\hbox{$\st\rm o$}\,}%
\def\otto{\,{\kern-1.truept\leftarrow\kern-5.truept\to\kern-1.truept}\,}
\def\tende#1{\,\vtop{\ialign{##\crcr\rightarrowfill\crcr
 \noalign{\kern-1pt\nointerlineskip} \hskip3.pt${\scriptstyle
 #1}$\hskip3.pt\crcr}}\,}

\newbox\strutboxa
\setbox\strutboxa=\hbox{\vrule height8.5pt depth2.25pt width0pt}
\def\struta{\relax\ifmmode\copy\strutboxa\else\unhcopy\strutboxa\fi}
\def\W#1{#1_{\kern-3pt\lower7.5pt\hbox{$\widetilde{}$}}\kern2pt\,\struta}
\def\T#1{{#1_{\kern-3pt\lower7pt\hbox{$\widetilde{}$}}\kern3pt}}
\def\VV#1{{\underline #1}_{\kern-3pt
\lower7pt\hbox{$\widetilde{}$}}\kern3pt\,}

\newdimen\xshift \newdimen\xwidth \newdimen\yshift \newdimen\ywidth

\def\ins#1#2#3{\vbox to0pt{\kern-#2\hbox{\kern#1 #3}\vss}\nointerlineskip}

\def\eqfig#1#2#3#4#5{%
\par\xwidth=#1\xshift=\hsize\advance\xshift%
by-\xwidth\divide\xshift by 2%
\yshift=#2\divide\yshift by 2%
{\hglue\xshift \vbox to #2{\vfil%
#3\includegraphics{#4.eps}%
}\hfill\raise\yshift\hbox{#5}}}%

\newcommand\revtex{{R\kern-0.4mm\lower0.5mm\hbox{E}\kern-0.4mm V\kern-0.3mm%
\lower0.5mm\hbox{T}\kern-0.4mm E\kern-.3mm \lower0.5mm\hbox{X}}}
\fiat

\begin{document}

\centerline{\titolone Microscopic chaos and
%}
%
%\centerline{\titolone 
macroscopic entropy in fluids}
\*
\centerline{\bf Giovanni Gallavotti}

\centerline{Fisica and I.N.F.N. Roma 1}
\centerline{
%7 July 2006 ;
\today}

\*\*

\0{\bf Abstract: \it In nonequilibrium thermodynamics ma\-cro\-scopic
entropy creation plays an important role. Here we study, from various
viewpoints, its relation with the phase space contraction, which has
been recently proposed as an apparently alternative quantity.}  \* \*

\0{\it Keywords: \sl Entropy, Nonequilibrium Thermodynamics, Fluids,
Navier-Stokes, Thermostats, Chaotic hypothesis} \*

\0{\bf 1. \it Thermostats, entropy creation and phase space contraction}
\*

Studying stationary states of mechanical systems in interaction with
thermostats the latter are often modeled by systems of particles
subject to anholonomic constraints. The equations of motion take the
form $\dot x=f_{\V E}(x)$ where $x$ is a point in phase space and $\V
E$ are parameters controlling the size of the acting nonconservative
forces.

It has appeared natural to define {\it entropy creation rate} as the
divergence $\s(x)\defi-\sum_j \dpr_{x_j} f_{\V E,j}(x)$.  There are
other natural definitions of entropy creation rate and here we study
their relation with the above phase space divergence. The aim is to
find such a relationship for a system that can be considered to be
described by a continuum evolving under macroscopic equations.

\eqfig{110pt}{90pt}{}{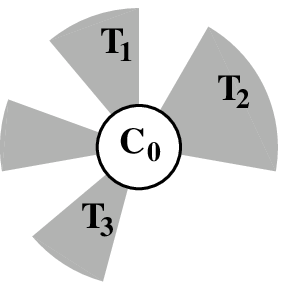}{Fig1}

\0{\nota Fig.1 Reservoirs occupy finite regions outside $\CC_0$,
{\it e.g.} sectors $\CC_i\subset R^3$, $i=1,2\ldots$. Their particles are
constrained to have a {\it total} kinetic energy $K_i$ constant, by
suitable forces, so that the reservoirs ``temperatures'' $T_i$ are
well defined.\vfil} 
\*

To be concrete and in rather ample generality we imagine a system
$\CC_0$ of particles enclosed in a container, also called $\CC_0$,
with elastic boundary conditions surrounded by a few thermostats which
consist of particles, all of unit mass for simplicity, interacting
with the system via short range interactions, through a portion
$\dpr_i{\CC_0}$ of the surface of ${\CC_0}$, and subject to the
constraint that the total kinetic energy of the $N_i$ particles in the
$i$-th thermostat is $K_i=\fra12 \dot{\V X}_i^2=\fra32 N_i k_B T_i$.
A symbolic illustration is in Fig.1. The equations of motion will be

$$
\ddot{\V X}_0=-\dpr_{\V X_0}\Big( U_0(\V X_0)+\sum_{j>0}
W_{0,j}(\V X_{0},\V X_j)\Big)+\V E(\V X_0),
$$
\begin{equation}\ddot{\V X}_i=-\dpr_{\V X_i}\Big( U_i(\V X_i)+
W_{0,i}(\V X_{i},\V X_j)\Big)-\a_i \dot{\V X}_i\hbox{\hglue1.1truecm}\label{e1}
\end{equation}
with $\a_i$ such that $K_i$ is a constant. Here $W_{0,i}$ is the
interaction potential between particles in $\CC_i$ and in $\CC_0$,
while $U_0,U_i$ are the internal energies of the particles in
$\CC_0,\CC_i$ respectively. We imagine that the energies $W_{0,j},U_j$
are due to {\it smooth} translation invariant pair
potentials; repulsion from the boundaries of the containers will be
elastic reflection. It is assumed, in Eq.(\ref{e1}) that there is no
direct interaction between different thermostats: their particles
interact directly only with the ones in $\CC_0$.  Here $\V E({\V X}_0)$
denotes possibly present external positional forces stirring the
particles in $\CC_0$.

Since the work per unit time that particles outside the thermostat
$\CC_i$ (hence in $\CC_0$) exercise on the particles in it, is
$Q_i\defi-\dpr_{{\V X}_{i}}W_{0,i}(\V X_{0},{\V X}_i)\cdot\dot{\V X}_i$ and it
can be interpreted as the ``amount of heat $Q_i$ entering'' the
thermostat $\CC_i$, energy conservation yields

\begin{eqnarray}\fra{d}{dt} \big(\fra{1}2\dot{\V X}_i^{2}+ U_i)\=\dot U_i=-
\a_i \dot{\V X_i}^2
+Q_i\label{e2}\end{eqnarray}
and the contraints on the thermostats kinetic energies give
$\a_i\=\fra{Q_i-\dot U_i}{3N_i k_B T_i}$.

Set $x=(\V X_i,\dot{\V X}_i)_{i=0,..}$ and write Eq.(\ref{e1}) as $\dot x=
f_{\V E}(x)$. The divergence $\s(x)=-\sum_j\dpr_{x_j}f_{\V E,j}(x)$ of
the equations of motion in phase space is readily computed from
Eq.(\ref{e1}), see also \cite{Ga06}, and is $\s(x)={\sum_{i>0}}
\fra{Q_i}{k_B T_i} +\dot R$ or 

\begin{eqnarray}
\s(x)=\e(x)+\dot R,\kern1truecm& \hbox{with}\nonumber\\
\e(x)=\dt\sum_{i>0} \fra{Q_i}{k_B T_i}\hbox{\hglue1.2truecm}
\label{e3}\end{eqnarray}
where $R=\sum_{i>0} \fra{U_{i}}{k_B T_i}$: here and $\fra{Q_i}{k_B
T_i},\fra{U_{i}}{k_B T_i}$ should really be multiplied by $(1-\fra1{3
N_i})$: this neglection of $O(N_i^{-1})$ is made just to simplify the
formulae as, in any event, we shall be interested in cases in which
$N_0,N_i\gg1$.
\*

\0{\it Remark:} (i) $\e(x)\defi{\sum_{i>0}} \fra{Q_i}{k_B T_i}$
can be called naturally the {\it entropy creation rate} and, therefore, 
Eq.(\ref{e3}) has a physical meaning: {\it entropy is created at the
boundary of the system}. Creation {\it
really} takes place where the walls get in contact with the
thermostats, where the temperatures $T_i$ are defined.  
\\
(ii) Note that if particles in $\CC_0$ were {\it also} subject to an
isokinetic constraint $\fra{1}2(\dot{\V X}_0)^2=\fra32 N_0 k_B T_0$  
phase space contraction would simply be changed by the addition of
$\fra {Q_0}{k_B T_0}$ with $Q_0$ being the work done per unit time by
the thermostats in $\CC_i$, $i>0$, on particles in $\CC_0$; also
$R$ will contain an extra term proportional to $\dot U_0$.
\\
(iii) The divergence $\s(x)$ is {\it different} from 
the entropy creation rate $\e(x)$. Their difference is  a
``total time derivative'', see  Eq.(\ref{e3}), therefore there is a relation
between the time averages
$a\defi \fra1\t\ig_{-\fra\t2}^{\fra{\t}2} \s(S_t x)\,dt$
and $a_0\defi \fra1\t\ig_{-\fra\t2}^{\fra{\t}2} \e(S_t x)\,dt$, namely

\begin{eqnarray}
a=a_0+\fra1\t\big(R(S_{\fra\t2}x)-R(S_{-\fra\t2}x)\big).
\label{e4}\end{eqnarray}
This means that the observables $a$ and $a_0$ will have the {\it same}
distribution with respect to any stationary distribution in the limit
$\t\to\io$ if $R$ is a bounded function (as in our case). More general
and ``singular'' interaction potentials could be considered and would
lead to essentially equivalent conclusions, following the results in
\cite{BGGZ05}.  
\\ 
(iv) Note that also phase space contraction of a system in contact
with isokinetic thermostats has a precise physical meaning as it {\it
equals minus the sum of the dimensionless free energy creation rates}
$-\fra{\dot U_i}{k_B T_i}+\fra{Q_i}{k_B T_i}$ of the thermostats.

\* 
\0{\bf 2. \it Macroscopic fluids}
\*

The above analysis shows that the two notions of entropy creation rate
$\e(x)$ in Eq.(\ref{e3}) and of phase space contraction $\s(x)$ are
{\it related but different}. 

They have {\it in particular} the same stationary average, as they
differ by a total derivative $\dot R$. As a consequence, not only the
averages of $\s$ and $\e$ are equal but also that the fluctuations of
their finite time averages, {\it i.e.} of $a$ and $a_0$ in
Eq.(\ref{e4}), are the same in the limit $\t\to\io$: so that
properties known for the fluctuations of $\s$ imply corresponding
properties for the fluctuations of the physically meaningful entropy
creation rate $\e$. This is relevant because, in the literature,
several results have been derived concerning the fluctuations of the
time averages of the phase space contraction, see \cite{Ga02}.

Therefore it is of some interest to see what the above mechanical
notion of entropy creation rate becomes in a system which can be
considered as a continuum in a stationary state and in local
equilibrium. In fact for such a system an independent definition of
entropy creation is classical, \cite{DGM84}. We check that the two
notions coincide up to a total derivative.

Consider, in $\CC_0$, a system of particles which can be regarded as a
Navier-Stokes (NS) continuum in a stationary state and in contact with
fixed walls on which, at each boundary point $\Bx\in\dpr \CC_0$,
temperature is prescribed at a value $T(\Bx)$ because the surface
element $d s_\Bx$ is in contact with a thermostat (as idealized in
Fig.1) whose internal potential energy per unit surface is
$U_{ext}(\Bx)$. Then the entropy creation rate according to
Eq.(\ref{e3}) will be

\begin{eqnarray}\e=\ig_{\dpr\CC_0}\fra{Q(\Bx)}{k_B T(\Bx)} 
ds_{\Bx},\label{e5}\end{eqnarray}
where $Q(\Bx)$ is the amount of work per unit time and unit surface
that the fluid performs on the thermostat in contact with the surface
element $ds_\Bx$, while phase space contraction $\s$ will differ from
this by $-\fra{d}{dt}\ig_{\dpr\CC_0}\fra{U_{ext}(\Bx)}{k_B
T(\Bx)}\,ds_\Bx$.

There are no complete derivations of the NS equations from molecular
models: however all attempts (which achieve the result under
reasonable extra assumptions) deal with limiting regimes implying
restrictions on initial data and involving a length scaling of
$O(\d^{-1})$, a time scaling of $O(\d^{-2})$, (hence) a velocity
scaling $O(\d)$ and become exact in the limit as $\d\to0$ ($\d$ being
a dimensionless scaling parameter).

Here we shall assume that the NS equations can be also obtained from a
molecular model under a suitable scaling of space and time variables.
We shall therefore consider microscopic initial data with velocity and
position distributions with average fields (of density $\r(\V x)$, of
kinetic energy, {\it i.e.}  temperature, $T(\V x)$ and velocity $\V
u(\V x)$) consistent with initial values corresponding to a
continuum. And we shall suppose that they evolve so that average
velocity, density, kinetic energy satisfy NS with good approximation,
and exactly in the limit in which some scaling parameter
$\d\to0$. The temperature field is identified with the average
kinetic energy per particle, with velocities measured with respect to
the average velocity $\V u$.

Physically $\d$ is a parameter measuring ``how far from a continuum
the microscopic structure is''; it can be identified with the ratio
between the molecular free path and the length scale of the variation
of the macroscopic velocity and temperature fields.

Therefore in the limit $\d\to0$ each volume element will contain an
infinite number of particles and fluctuations will be
suppressed; however the {\it average} entropy creation will be defined
and, by Eq.(\ref{e3}), be

\begin{eqnarray}\media{\e}_\m=-\ig_{\dpr\CC_0}\k\, 
\fra{\V n(\Bx)\cdot\V\dpr\, T(\Bx)}{k_B T(\Bx)} 
ds_{\Bx}\label{e6}\end{eqnarray}
where $\k$ is the thermal conductivity, $\V n$ is the outer normal to
$\dpr\CC_0$ and {\it the average is intended over a time scale long
compared to the microscopic time evolution but macroscopically short}.

The subscript $\m$ indicates that the time average $\media{\e}_\m$ of
Eq.(\ref{e5}) coincides with the average with respect to a probability
distribution $\m$ that is defined on the microscopic phase space and
that we assume to be identified with a kind of evolving SRB
distribution representing local stationarity. {\it I.e.} we assume the
validity of a ``Chaotic hypothesis'', \cite{Ga06}, not only globally
and asymptotically in time but also locally and over times long
compared to the microscopic time evolution but macroscopically
short. This means that $\m$ may change with time over macroscopic time
scales and is, therefore, in general different from the true SRB
distribution that would describe the system over time scales very long
with respect to macroscopic times.

Hence suppression of fluctuations will not mean that the
$\m$-averages, over times long with respect to microscopic scales,
defining $\V u(\V x), T(\V x)$ or the average $U(\V x)\defi
\media{U}_\m(\V x), \V x\in \CC_0$, of the microscopic energy density
$U$ or the average energy per unit surface of the external thermostats 
$U_{ext}(\Bx)\defi$$\media{U_{ext}}_\m(\Bx), \Bx\in\dpr \CC_0$, over
such scales will not continue to vary over macroscopic times.

In other words we are saying that there are two basic time scales: at
times long with respect to the microcopic time scale (determined by
the molecular free path) the volume elements of the fluid are in a
stationary state and the distribution $\m$ used to compute the
averages can be thought as being locally stationary on times short
compared to the macroscopic times. As time evolves on the macroscopic
time scale the distribution $\m$ also changes: to average, asymptotically,
to the true SRB distribution for the global stationary state.
 
Eq.(\ref{e6}) is the expression corresponding to Eq.(\ref{e3}) derived
from molecular dynamics and it must be compared, for compatibility,
with the familiar expressions for the entropy creation rate in systems
described by macrosopic continua equations, \cite{DGM84}.

We consider here a viscous and thermally conducting NS-fluid in local
equilibrium.  Thus the local equilibrium entropy density $s$ depends
on the local temperature and density: $s=s(T,\r)$. Let $\VV\t'$ be the
stress tensor $\t'_{ij}=(\dpr_i u_j+\dpr_j u_i)$ in terms of the
velocity field $\V u$, $\h$ be the dynamical viscosity and $U(\V
x)=\media{U}_\m(\V x)$ be the internal energy density (sum of the
potential energy density and the kinetic energy density evaluated with
respect to the average velocity $\V u$). Then the NS equation are,
\cite[p.2,3,6,18]{Ga02},

\begin{eqnarray}
(1)\kern0.3truecm&\dpr_t\r+\V\dpr\cdot(\r\V u)=0\nonumber\\
(2)\kern0.3truecm&
\dpr_t\V u+\W u\cdot\W\dpr\, \V u=-{1\over\r}\V\dpr\, p
   +\fra{\h}\r \D\V u+\V g\label{e7}\\
(3)\kern0.3truecm&\dpr_t U+\V \dpr\cdot(\V u U)=\h\,\VV\t'\, 
\V\dpr \,\W u+\k\D T-p\,\V\dpr\cdot\V  u\nonumber\\
(4)\kern0.3truecm&T\,(\dpr_t s+\V \dpr\cdot(\V u s))\,=\,\h\,
   \VV\t'\, \V\dpr \,\W u+\k\D T\nonumber
\end{eqnarray}
here $\V g$ is a (nonconservative) external force generating the fluid
motion and $p$ is the physical pressure.  The conditions at the
boundary of the fluid container ${\CC_0}$ will be time independent,
$T=T(\Bx)$ and $\V u=\V0$ (no slip boundary).

As mentioned, Eq.(\ref{e7}) are macroscopic equations that can be
valid only in some limiting regime. Given a system of particles with
short range pair interactions let $\d$ be the mentioned natural
dimensionless scaling parameter; then a typical conjecture is: for
suitably restricted and close to local equilibrium initial data (see
\cite[p.21]{Ga02} for examples) {\it on time scales of $O(\d^{-2})$
and space scales $O(\d^{-1})$ the evolution of $\r,\V u,T,U,s$ follows
the incompressible NS equation}, \cite[p.30]{Ga02}.

The classical entropy creation rate in nonequilibrium
thermodynamics of an {\it incompressible fluid} is, \cite[p.6]{Ga02},

\begin{eqnarray}k_B \e_{classic}=\ig_{\CC_0}\Big(\k\, 
\big(\fra{\V\dpr T}{T}\big)^2
+\h\, \fra1T{\VV\t'\,\V\dpr \W u}\Big)\,d\V x.\label{e8}\end{eqnarray}
By integration by parts and use of the first and
fourth of Eq.(\ref{e7}), $k_B \e_{classic}$ becomes, if $S\defi
\ig_{\CC_0}
s\,d\V x$ is the total thermodynamic entropy of the fluid,

\begin{eqnarray}
&\dt\ig_{\CC_0}\Big(-\k\,\V\dpr T\,\cdot\,\V\dpr T^{-1}
+\h \,\fra1T{\VV\t'\,\V\dpr \W u}\Big)\,d\V x=\hbox{\hglue1.7truecm}\nonumber\\
&\dt=
-\ig_{\dpr {\CC_0}} \k\, \fra{\V n\cdot\V\dpr T}T \,ds_\Bx+
\ig_{\CC_0}\fra{(\k\D T+\h\, \VV\t'\V\dpr\W u)}T d\V x=   \label{e9}\\
%\noalign{\vglue.05mm}
&\dt=
-\ig_{\dpr {\CC_0}} \k \,\fra{\V n\cdot\V\dpr T}T\,ds_\Bx+
\dot S+\ig_{\CC_0} 
\V u\cdot\V\dpr s\,d\V x=\hbox{\hglue1.2truecm}   \nonumber      
\\
&\dt=
-\ig_{\dpr {\CC_0}}\k\, \fra{ \V\dpr T\cdot\V
  n}T\,ds_\Bx+\dot S\hbox{\hglue4.3truecm}\nonumber
\end{eqnarray}
{\it i.e.} this {\it still leads to} the expression Eq.(\ref{e6}),
``local on the boundary'' or ``localized at the contact between system
and thermostats'', since $\V u\cdot\V n\=0$ by the boundary
conditions, {\it plus the time derivative of the total
``thermodynamic entropy'' of the fluid}.  \*

\0{\it Remarks:} (i) An identical analysis can be performed for
{\it Rayleigh's convection model}, widely used to test ideas on
turbulence since \cite{Lo63}: the result is the same because the extra
term that would appear in Eq.(\ref{e9}), see \cite[p. 47]{Ga02}, would be
proportional to $\ig_{\CC_0} u_z d\V x$ which vanishes because the
motion has no net momentum in the $z$ direction.
\\
(ii) It should be noted that in the limit $\d\to0$, {\it i.e.} when the NS
equations are expected to become rigorously exact, the 
Eq.(\ref{e8}) simplifies: only the first term in {\it r.h.s.} remains
because the velocity $\V u$ scales as $O(\d)$, \cite[p.26]{Ga02}.

\* 
\0{\bf 3. \it Incompressible continua viewpoint}
\*

The above analysis leads to a further natural question: whether the
phase space contraction and the entropy creation rate can be computed
if we consider an incompressible fluid {\it in local equilibrium and
observed on a short but macroscopic time scale} and imagine, as it is
tempting to do and often done, \cite{DGM84,Ga02}, its small
macroscopic volume elements as a collection of small thermostats in
contact with reservoirs consisting in the neighboring volume elements:
if we apply the general theory of Sec.1 would the results be
consistent with the ones in Sec.2?

In establishing a comparison it should, however, be noted that the
volume elements $E=d\V x$ are not separated by walls, hence they can
exchange particles, and they also move.  In order to be able to treat
volume elements as systems on their own we imagine that their size is
$\l$ with $\l$ macroscopically small but microscopically large:
certainly such length scale $\l$ is $\ll L\sim\big(\fra1T\fra{\dpr
T}{\dpr x}\big)^{-1}$ where $L$ is the macroscopic scale of the
container ${\CC_0}$.

Furthermore we have to assume that molecules diffuse, in a
characteristic evolution time, over a distance $\ll \l$. The diffusion
coefficient is $D= O(\fra{k_B T}{m r^2 \r v})$ with $v$ the average
speed, $v=O(v_{sound})$, $m$ the mass of the molecules, $r$ their
radius and $\r$ the numerical density, and a characteristic time scale
is $\th=\fra{m D}{k_B T}$. The distance traveled by diffusion in
the latter time scale is $(D\th)^{\fra12}$ ($\sim 10^{-2}$cm in
air at normal conditions).  

We assume, as above, that the local quantities, velocity field
and temperature field, $\V u(\V x),T(\V x)$, as well as the energy
fields $\media{U}_\m(\V x)=U(\V x), \V x\in \CC_0$, and
$\media{U_{ext}}_\m(\Bx)=U_{ext}(\Bx), \Bx\in \dpr\CC_0$, evolve on
a time scale much slower than the microscopic time scale $\th$ and can
be considered constant on the latter time scale. The expression
Eq.(\ref{e8}) should be regarded as an average over a long microscopic
time (but over a short macroscopic time).

Suppose that the conditions allow us to consider a volume element in
a fluid as a thermostated system in a stationary state in contact with
thermostats made of the neighboring elements. {\it I.e.\ } suppose
that the quantity $\d$ introduced after Eq.(\ref{e7}) is small and the
diffusion across the elements boundaries is not important enough to
make the identity of the volume elements ill defined ({\it i.e.}
$\d\ll\l\ll L$). Then we can apply the analysis leading from
Eq.(\ref{e1}) to Eq.(\ref{e3}) and conclude that {\it up to a total
derivative $\dot R$} the phase space contraction of the total system,
{\it i.e.} fluid plus thermostats, is (see also remark (ii) in Sec1)

\begin{eqnarray}\e(x)=\sum_E \sum_{E'} \fra{Q_{E,E'}}{k_BT_{E'}}
\label{e10}\end{eqnarray}
where $Q_{E,E'}$ is the amount of work that the particles in a given
volume element $E$ perform over the neighboring elements $E'$, see
Eq.(\ref{e3}). The quantity $\e(x)$ will be the entropy creation rate
defined by regarding the volume elements as small thermostats in
stationary state.

If the average heat current is $-\k\, \V\dpr T$ and the element $E$ is
imagined with the bases orthogonal to the gradient of $T$, the average
contribution to $Q_{E,E'}$ for $E'$ adjacent to the upper base of $E$
is $-\k \,\V\dpr T\cdot \V n\, \l^2$ ($\l^2$ = area of the base) and it is
opposite to the contribution from the lower base; therefore the
quantity $\sum_{E'} \fra{Q_{E,E'}}{T_{E'}}$ has average 

\begin{eqnarray} -\k\, \V\dpr T\cdot \V n \,\l^2\,(\fra1{T_+}-\fra1{T_-})=-\k\,
\V\dpr T \cdot \V\dpr T^{-1}\l^3,\label{e11}\end{eqnarray}
if $T_\pm$ are the temperatures at the two bases; hence summing
over $E$: $k_B\media{\e}_\m=\ig_{\CC_0}
\V\dpr(\fra1{T})\cdot(-\k\,\V\dpr T) \,d {\V x}$ which can be written
in the more common and familiar form $\k\,\ig_{\CC_0} \big(\fra{\V\dpr T}
T\big)^2d\V x$. Then if $\dpr_t T=0,\D T=0$ and $\V u=\V 0$
Eq.(\ref{e8}), therefore Eq.(\ref{e6}), follows by partial integration.

More generally in presence of time dependence and non vanishing
velocity field there will be an extra amount of energy transfered to
elements adjacent to $E$ and due to diffusion across the bases: it can
be evaluated in the same way as above to be $\h\, \VV\t'\cdot \V
n\,(\W u_+-\W u_-)\,\l^2$ if $\VV\t'$ is the stress tensor and it
changes $k_B\media{\e}_\m$ to Eq.(\ref{e8}), {\it hence to
Eq.(\ref{e6}) plus $\dot S$} (see Eq.(\ref{e9})) if account is taken
of the fourth of Eq.(\ref{e7}).  

The above viewpoint therefore also leads to the classical expression for
the entropy creation rate.  \*

\0{\bf 4. \it Phase space contraction}
\*

Finally it is interesting to remark that not only the entropy creation
rate but also the phase space contraction can be computed along the
above lines for macroscopic continua. Regarding each volume element
$E$ as a thermostated system in a stationary state with a fixed
temperature, the average phase space contraction is deduced from
Eq.(\ref{e3}) taking into account the $\dot R$ contribution.

Hence, setting $k_B\,e(\V x)\defi \Big(\k\, \big(\fra{\V\dpr
T}{T}\big)^2 +\h\, \fra1T{\VV\t'\,\V\dpr \W u}\Big)$, the average
phase space contraction is $\ig_{\CC_0}\big({e(\V x)}-\fra{\dot U(\V
x)} {k_B T(\V x)}\big)\,d\V x-\ig_{\dpr\CC_0 }\fra{\dot
U_{ext}(\Bx)}{k_B T(\Bx)} \,ds_{\Bx}$, where $U(\V x)=\media{U}_\m(\V
x)$ denotes the average potential energy density while with
$U_{ext}(\Bx)=\media{U_{ext}}_\m(\Bx)$ we denote the average potential
energy density per unit surface of the thermostats.

Remark that the time derivative of the potential energy density $\dot
U(\V x)$ could be replaced by the time derivative of the total energy
density at $\V x$: in fact the kinetic energies do not contribute to
the total derivative of the energy density since they have been
supposed to be constant because each volume element is regarded to
have a constant kinetic energy, {\it i.e.} a well defined temperature,
and in the above relations the energies appear through their time
derivative). Hence energy conservation, $\dot U(\V x)=
\h\,\VV\t'\cdot\V\dpr\,\W u+\k\,\D T$, see (3) in Eq.(\ref{e7}), for
incompressible fluids and partial integration of the contribution
$\k\big(\fra{\V\dpr T}T\big)^2$ to $e(\V x)$, see Eq.(\ref{e8}), in
the integral $\ig_{\CC_0}\big(e(\V x)-\fra{\dot U(\V x)}{k_B T(\V
x)}\big)\,d\V x$ leave us with a boundary term {\it which is just
Eq.(\ref{e6}) minus} $\ig_{\dpr \CC_0}\fra{\dot U_{ext}(\Bx)}{k_B
T(\Bx)}ds_{\Bx}$ as it could be guessed from the expression for $\s$
preceding Eq.(\ref{e3}).

Therefore even if we regard the fluid volume elements as thermostated
systems in stationary nonequilibrium we are led to the expected
relation between average entropy creation $\media{\e}_\m$ and average
phase space divergence $\media{\s}_\m$, namely (if $x$ denotes the
fields determining the state of the fluid)

\begin{eqnarray}\media{\s}_\m=\media{\e}_\m-\ig_{\dpr \CC_0}
\fra{\dot U_{ext}(\Bx)}{k_B T(\Bx)}ds_{\Bx}\label{e12}.\end{eqnarray}

Analysis of compressible fluids, unfortunately more difficult,
should also be attempted: the first difficulty will be, of course, that
it is not clear under which scaling the compressible NS equations
should hold as a reasonable approximation.
\*

\0{\bf 5. \it Conclusion}
\*

The conclusion is that the {\it entropy creation rate} evaluated from the
microscopic model, Eq.(\ref{e1}), and the microscopic definition of
heat ceded to a thermostat (see Eq.(\ref{e2})) 

\0(1) differs ``by a total derivative'', $\dot R$, from the
microscopic phase space contraction (see Eq.(\ref{e3})), 
\\ 
(2) if the system can be regarded as a NS continuum in stationary
local equilibrium, its average coincides with the classical entropy
creation rate up to a total derivative, $\dot S$, see
Eq.(\ref{e9}),
\\
(3) in the latter situation its average coincides with the entropy
creation obtained by regarding the continuum as constituted by volume
elements each of which is a thermostat in a stationary state
exchanging heat with the neighboring volume elements, see
Eq.(\ref{e8}), and differs by a total derivative from the average
phase space contraction evaluated by regarding the continuum in the
same way, see Eq.(\ref{e12}).

\* \0{\bf Acknowledgements: \rm I am
indebted to A.Giuliani, F.Zamponi and, in particular, to F.Bonetto for
important suggestions.}

%\*\0\revtex
%\vfill\eject

\nota 
%\bibliography{0Bibcaos} 
%\bibliographystyle{apsrev}
\bibliographystyle{unsrt}

\*

\0e-mail: {\tt giovanni.gallavotti@roma1.infn.it}
\\
web: {\tt http://ipparco.roma1.infn.it}
\*\0\revtex
\end{document}